\documentclass[aps,graphicx,prb,twocolumn,groupedaddress,showpacs,notitlepage]{revtex4-1}
\usepackage{graphicx}
\usepackage{amsmath,amssymb}
\usepackage{color}

\newcommand{\bra}[1]{\left<#1\right|}
\newcommand{\ket}[1]{\left|#1\right>}

\newcommand{\abs}[1]{\left|#1\right|}

\begin{document}
\title{Strong Three-magnon Scattering in Cuprates by Resonant X-rays}
\author{Luuk J. P. Ament and Jeroen van den Brink}
\affiliation{Institute-Lorentz for Theoretical Physics, Universiteit Leiden, 2300 RA Leiden, The Netherlands\\
Institute for Theoretical Solid State Physics, IFW Dresden, 01171 Dresden, Germany}
\date{\today}

\begin{abstract}
We show that Resonant Inelastic X-ray scattering (RIXS) is sensitive to three-magnon excitations in cuprates.   Even if  it requires three electrons to simultaneously flip their spin, the RIXS tri-magnon scattering amplitude is not small. At the Cu $L$-edge its intensity is generally larger than the bi-magnon one and at low transferred momentum even larger than the single-magnon intensity. At the copper $M$-edge the situation is yet more extreme: in this case three-magnon scattering is dominating over all other magnetic channels.  

\end{abstract}


\maketitle

{\it Introduction.} Traditionally, inelastic neutron scattering is the technique of choice to measure the dispersion of elementary magnetic excitations in magnetic materials. Only very recently photon scattering became competitive in this respect, with Resonant Inelastic X-ray Scattering (RIXS)\cite{Schuelke2007,Kotani2001} getting energy-sensitive enough to observe magnon excitations~\cite{Hill2008,Braicovich2009a,Schlappa2009,Ghiringhelli2009,Braicovich2009b,Freelon2008} -- an experimental endeavor underpinned by theory\cite{Brink2007,Nagao2007,Forte2008,Ament2009,Haverkort2009}. RIXS is a photon in - photon out technique in which the energy of the incoming x-ray photon is tuned to an absorption edge of the material under study.
By tuning to the copper $L_3$-edge, for instance, the {\it single-magnon} dispersion in La$_2$CuO$_4$ was recently mapped out and found to be in perfect agreement with the neutron scattering data\cite{Braicovich2009b}. At the Cu $K$-edge the RIXS process is indirect~\cite{Brink2006} and single-magnon scattering is forbidden. Instead, {\it bi-magnon} creation is allowed\cite{Brink2007,Nagao2007,Forte2008} and indeed observed\cite{Hill2008}. In the latter case magnetic RIXS is thus complementary to optical Raman scattering, which also measures the bi-magnon, but at zero momentum transfer only\cite{Devereaux2007}.
 
Here we show that in addition to these rather familiar magnetic modes, RIXS also has the unique potential to examine properties of a virtually unexplored magnetic excitation that cannot be probed directly by any other technique: the {\it tri-magnon}.  Tri-magnons are three-magnon states that have been identified as important and generic features of gapful, low-dimensional quantum spin systems\cite{Els1997}. Theoretically it has long been known that three reversed spins can form a tri-magnon boundstate\cite{Millet1974,Himbergen1976,Cyr1996}, but the spectroscopic means to directly create and probe such an excitation has been lacking so far. We find that in RIXS the tri-magnon scattering amplitude is not small, even if it requires three electrons to simultaneously flip their spin. At the Cu $L$-edge the three-magnon intensity is generally larger than the bi-magnon one and at low transferred momentum $\bf q$ we find that it is even larger than single-magnon intensity. At the copper $M$-edge the situation is yet more extreme, as in this case the three-magnon scattering channel is the dominant one.  It is thus clear that for a quantitative interpretation of magnetic $L$- and $M$-edge RIXS data three-magnon scattering is indispensable. At the same time the presence of a substantial tri-magnon scattering amplitude illustrates that RIXS has an unprecedented potential to measure properties of interesting multi-particle states in strongly correlated magnets.

{\it RIXS cross section.} In RIXS\cite{Kotani2001} an x-ray excites an electron from a core orbital to an empty state above the Fermi level. Rapidly the high-energy core-hole decays  again. In this process the electron filling the core-hole emits an x-ray photon, which is then detected. In the short-lived intermediate state the motion of the valence electrons is affected by the presence of the core-hole potential. In case of indirect RIXS the perturbing core-hole interaction drives the creation of elementary excitations in the solid. Direct RIXS has a different scattering mechanism: in that case the excited and de-excited electron simply involve different quantum states. The energy and momentum necessary to make excitations are lost by the scattered x-ray photons. By observing the energy loss $\omega$ and momentum loss ${\bf q}$ of the outgoing photons, one can thus determine the dispersion of the excited mode. 

The scattering amplitude $A_{fi}$ that takes an initial state $\ket{i}$ into a final state $\ket{f}$ is given by the Kramers-Heisenberg equation: $ A_{fi} = \sum_n \bra{f}\hat{D}\ket{n}\bra{n}\hat{D}\ket{i}/(E_i - E_n +  i\Gamma)$. The absorption of the incident photon and the emission of the outgoing photon are governed by the dipole operator $\hat{D} \propto {\bf A} \cdot {\bf p}$, with ${\bf A}$ the vector potential and ${\bf p}$ the momentum of the electron. The sum is over all intermediate states $\ket{n}$ with energies $E_n$ and inverse lifetime $\Gamma$. In RIXS experiments, the energy of the photon $E_i$ is tuned to a resonant edge: $E_i$ matches a set of $E_n$, which greatly enhances the scattering amplitude.

In general the evaluation of the Kramers-Heisenberg equation is a complex problem because of the nature of the intermediate states. In many cases, however, the intermediate states are very short-lived, and one can employ the lifetime $\hbar/\Gamma$ as a small parameter. This approach is formalized in the Ultra-short Core-hole Lifetime (UCL) expansion.\cite{Brink2005,Brink2006,Ament2007} The Kramers-Heisenberg relation is expanded in $H/\Gamma$, where $H$ is the Hamiltonian of the system including the core-hole. The result is $A_{fi} = \frac{1}{i\Gamma} \bra{f} \hat{D} \sum_{l=0}^{\infty}  \left[ (H-E_i)/i\Gamma \right]^l \hat{D} \ket{i}.$ Instead of evaluating the intermediate states, to first order one now only needs to compute the matrix elements of $H$.

{\it Single- and bi-magnon scattering.} The experimental effort to measure magnetic excitations with RIXS is focussed on compounds in the copper-oxide family\cite{Hill2008,Braicovich2009a,Braicovich2009b,Freelon2008,Schlappa2009}. To understand how three-magnon scattering comes about in these materials, we first briefly summarize the single- and bi-magnon scattering mechanism. Bi-magnon scattering at the Cu $K$-edge RIXS starts with an electronic $1s \rightarrow 4p$ transition, with an energy of around 8990 eV. In the intermediate state bi-magnon excitations are created dynamically because the $1s$ core-hole locally modifies the superexchange bonds\cite{Brink2007,Nagao2007,Forte2008}. Single-magnon excitations are forbidden because the total spin is conserved in the scattering processes. For undoped cuprates the magnetic Heisenberg Hamiltonian in the presence of a core-hole is\cite{Brink2007}
\begin{equation}
  H = J\sum_{i, \delta} {\bf S}_i\cdot {\bf S}_{i+\delta} - J^{\prime} \sum_i h^{\dag}_i h^{\phantom{\dag}}_i  \sum_{\delta} {\bf S}_i \cdot {\bf S}_{i+\delta} \label{eq:H},
\end{equation}
where we restrict ourselves for simplicity to nearest neighbor exchange between spins ${\bf S}_i$ at sites $i$, $\delta$ denotes the neighbors and  $h^{\dag}_i$ is the core-hole creation operator. The first term is the spin 1/2 Heisenberg model with exchange interaction $J$ and the second term corresponds to the perturbation of the magnetic bond strength due to the presence of a core-hole. In first order of the UCL expansion one immediately obtains the two-magnon scattering amplitude $A^{(2)}_{fi} ({\bf q}) \propto \bra{f} \sum_i e^{i{\bf q}\cdot {\bf R}_i} h^{\phantom{\dag}}_i H h^{\dag}_i \ket{i}$,where ${\bf R}_i$ is the position of the copper ion absorbing the incoming photon and ${\bf q}$ the transferred momentum.  It is important to note that in the limit ${\bf q} \rightarrow 0$ selection rules\cite{Forte2008} are such that the bi-magnon intensity rapidly vanishes $\propto {\bf q}^4$. 

In contrast, single-magnon scattering is allowed at the copper $L$- and $M$-edges\cite{Ament2009}, where the initial photo-excitation is $2p\rightarrow 3d$ ($L$-edge, around 930 eV) and $3p\rightarrow 3d$ ($M$-edge, around 75 eV). The reason is that the spin of the $p$ core-hole is not conserved in the intermediate states: it can flip via the spin-orbit coupling. The probability for spin-flip and non-flip are comparable since the core-hole spin-orbit coupling is very large, $\sim 20$ eV at the $L$-edge. Therefore, when the  core-hole is annihilated, a single-magnon can be created in the final state\cite{Ament2009,Haverkort2009}. The spin-flip amplitude for a single copper ion is $G({\boldsymbol \epsilon},{\boldsymbol \epsilon}') \approx \sum_{m_j} \bra{f} \hat{D} \ket{m_j} \bra{m_j} \hat{D} \ket{i}$ where $m_j$ labels either the $j=1/2$ or $j=3/2$ spin-orbit split intermediate states, depending on whether the $L_2$- or $L_3$-edge is used, respectively. $G({\boldsymbol \epsilon},{\boldsymbol \epsilon}')$ depends on the polarization vectors ${\boldsymbol \epsilon}$ and ${\boldsymbol \epsilon}'$ of the incoming and outgoing photons respectively. The resulting single-magnon scattering amplitude is $A^{(1)}_{fi} = -\frac{G({\boldsymbol \epsilon},{\boldsymbol  \epsilon}')}{\Gamma^2 N} \bra{f} \sum_i e^{i{\bf q}\cdot {\bf R}_i}  (S^+_i + S^-_i) \ket{i}$. In the ${\bf q} \rightarrow 0$ limit, not only the single-magnon energy vanishes, but also its spectral weight. Note that in addition to single-magnon scattering, at the $L$- and $M$-edges also bi-magnon scattering is allowed\cite{Braicovich2009a} because the intermediate state configuration is $3d^{10}$ and this singlet state completely blocks any superexchange: $J^{\prime}=J$ in Eq.~(\ref{eq:H}). But just as at the $K$-edge for ${\bf q} \rightarrow 0$  the bi-magnon intensity vanishes in leading order of the UCL also at the $M$- and $L$-edges. Consequently at the Cu $M$-edge single- and bi-magnon scattering have negligible intensities: due to the low photon energy at the $M$-edge the transferred momentum $|{\bf q}|$ is restricted to the center $10 \%$ of the Brillouin zone of a typical cuprate. In second order of the UCL\cite{Forte2008} a finite bi-magnon scattering amplitude appears at the $M$-edge, but as at this edge typically\cite{Yin1974,Coldea2001} $J/\Gamma \approx 0.1$, this channel is very weak too.

\begin{figure}
  \begin{center}
    \includegraphics[width=.9\columnwidth,clip,trim=3.5cm 2.6cm 3.5cm 3.6cm]{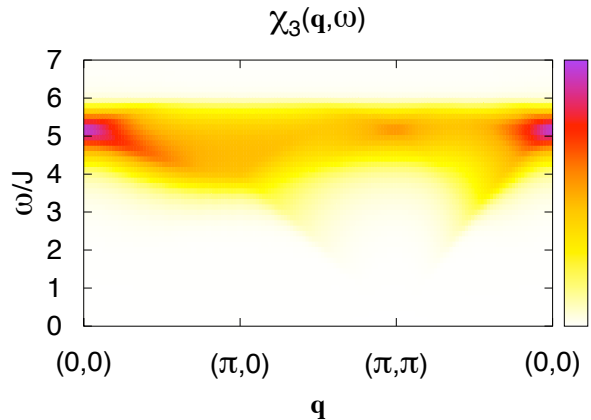}
    \caption{RIXS three-magnon response function $\chi_3 ({\bf q},\omega)$ for a two-dimensional Heisenberg model with nearest neighbor exchange.} 
    \label{fig:response}
  \end{center}
\end{figure}

{\it Three-magnon scattering mechanism.} 
Single- and bi-magnon excitations can be created independently of each other, but in principle they can also be combined. Such a flipped spin in the vicinity of a bi-magnon excitation can form a tri-magnon bound state. We will show that the RIXS selection rule that forbids single- and bi-magnon scattering at ${\bf q} = 0$ does not apply for three-magnon scattering. Therefore three-magnon scattering becomes, in terms of the UCL expansion, a first order scattering channel. Consequently three-spin modes give a large contribution to the magnetic RIXS spectra of the high-$T_c$ cuprates at the $M$- and $L$-edges for small transferred momenta. 

The three-magnon excitation mechanism is very local and a direct combination of the single-magnon and a bi-magnon creation mechanism outlined above. It is caused by the $3d^{10}$ intermediate state being screened by a bi-magnon excitation and then decaying via the spin-flip channel, which produces the third magnon.  Via the spin-flip decay channel this process contributes to the first order of the UCL expansion, with an amplitude $A^{(3)}_{fi} = -\frac{G({\boldsymbol \epsilon},{\boldsymbol   \epsilon}')}{\Gamma^2 N} \bra{f} \sum_i e^{i{\bf q}\cdot {\bf R}_i} (S^+_i + S^-_i) h^{\phantom{\dag}}_i H h^{\dag}_i \ket{i}$. Using the Hamiltonian in Eq.~(\ref{eq:H}) and integrating out the core-hole degree of freedom yields the three-magnon scattering amplitude
\begin{equation}  A^{(3)}_{fi} = \frac{J G({\boldsymbol \epsilon},{\boldsymbol
      \epsilon}')}{\Gamma^2 N} \sum_{i,\delta} e^{i{\bf q}\cdot {\bf
      R}_i} \bra{f} (S^+_i + S^-_i) {\bf S}_i \cdot {\bf S}_{i+\delta} \ket{i}.
\end{equation}
This scattering operator can be written in terms of magnon operators by expressing the spin operators in Holstein-Primakoff bosons and then Fourier and Bogliubov transforming it. With conventions as in Ref.~\onlinecite{Forte2008}, the three-magnon scattering amplitude then becomes
\begin{align}
A^{(3)}_{fi} &= -\frac{J z G({\boldsymbol \epsilon},{\boldsymbol  \epsilon}')}{2\Gamma^2 N^{3/2}} \sum_{{\bf k},{\bf p}} \left( u_{\bf k} - v_{\bf k} \right) \left[ \left( 1+\gamma_{{\bf k}+{\bf  q}} \right) u_{\bf p} v_{{\bf k}+{\bf p}+{\bf q}} \right. \nonumber \\
&\left. -\gamma_{{\bf k}+{\bf p}+{\bf q}} \left( u_{\bf p} u_{{\bf k}+{\bf  p}+{\bf q}} + v_{\bf p} v_{{\bf k}+{\bf p}+{\bf q}} \right) \right] \nonumber \\
&\left. \bra{f} \alpha^{\dag}_{\bf k} \alpha^{\dag}_{\bf p} \alpha^{\dag}_{-{\bf k}-{\bf p}-{\bf q}} \ket{i}.  \right.
\end{align}
The cross section at zero temperature is obtained by summing over all final states with three magnons: $I \propto \sum_f \abs{A_{fi}}^2 \delta (\omega - \omega_{fi})$ where $\omega_{fi}$ is the energy of the three-magnon final state $f$. We rewrite this as
$I \propto \frac{J^2 G({\boldsymbol \epsilon},{\boldsymbol \epsilon}')^2}{\Gamma^4 N} \chi_3 ({\bf q},\omega) $, where $\chi_3 ({\bf q},\omega)$ is the three-magnon response function.
 
 \begin{figure}
\begin{center}
\includegraphics[width=.39\columnwidth,clip,trim=6.5cm 2.0cm 7.8cm 3.6cm]{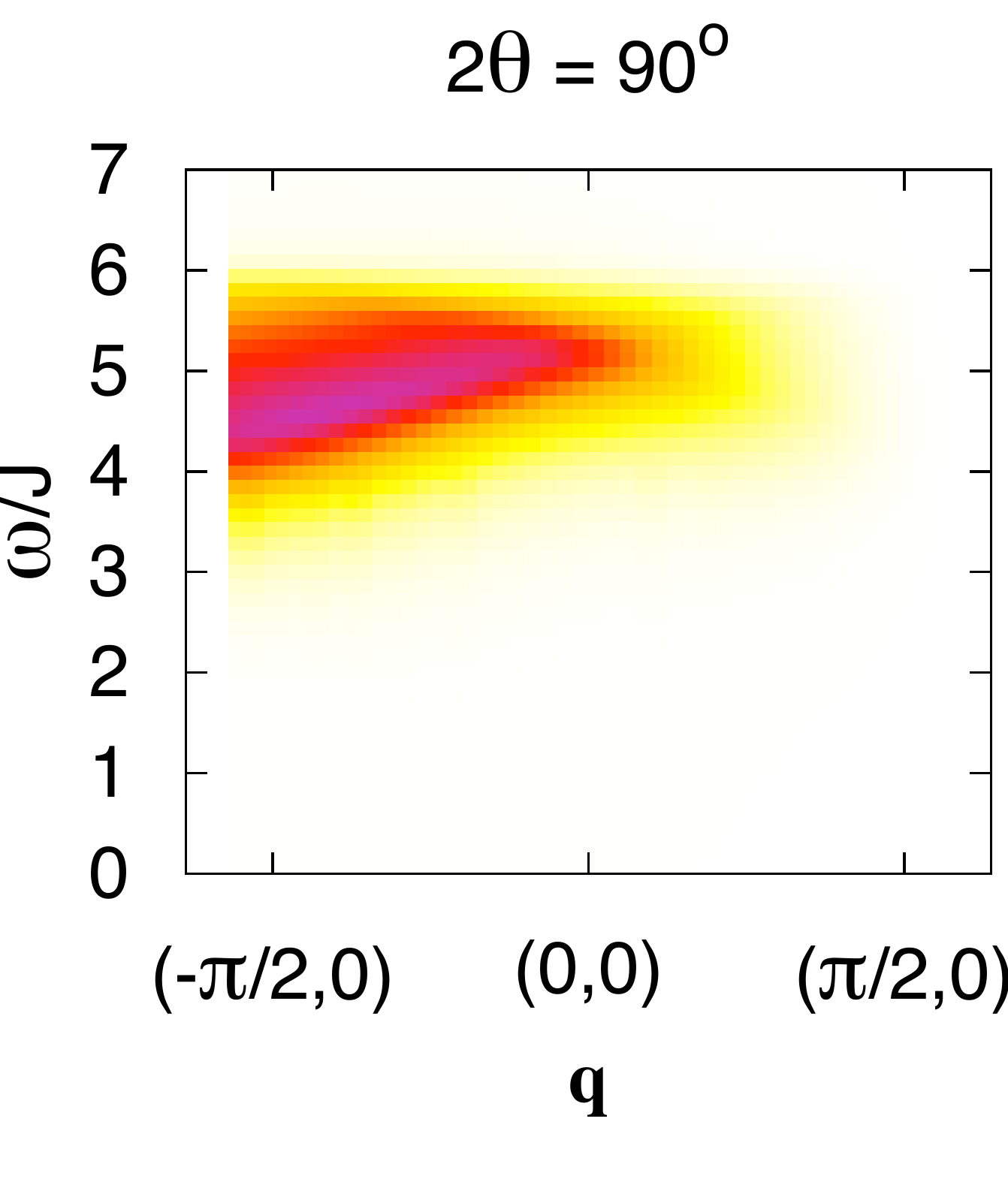}
\includegraphics[width=.59\columnwidth,clip,trim=4cm 2.1cm 3.5cm 3.7cm]{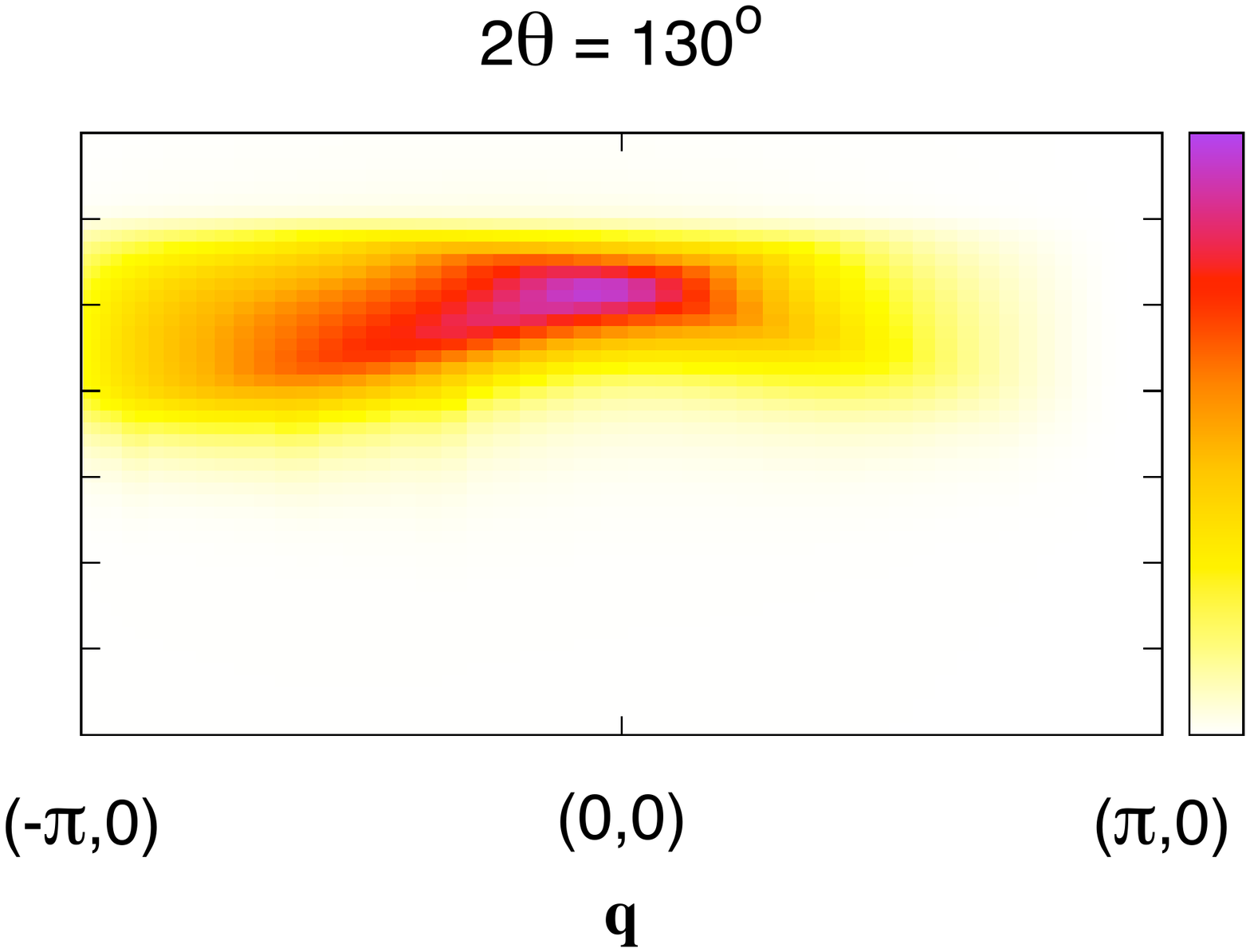}
\caption{Calculated three-magnon $L$-edge RIXS spectra for the case that the $[0,\pi]$-direction is perpendicular to the scattering plane with scattering angles $2\theta= 90^{\circ}$ and $130^{\circ}$ respectively. In $90^{\circ}$ geometry, not all momenta are experimentally accessible.}
\label{fig:Ledge}
\end{center}
\end{figure}

{\it Three-magnon spectral shape.} Fig.~\ref{fig:response} shows the three-magnon response function $\chi_3 ({\bf q},\omega)$ evaluated without magnon-magnon interactions. The three-magnon spectral weight is strongly peaked around  $\omega = 5.2J$ and at ${\bf q}=0$. The reason for this is two-fold. First, the three-magnon DOS, which is a convolution of many combinations of three magnon states whose total momentum adds up to ${\bf q}$, peaks close to its maximum energy of $3\times 2J$. It is in addition remarkable that the three-magnon response shows a strong ${\bf q}$ dependence, whereas the three-magnon DOS is quite featureless as the convolution tends to average out dispersions. The strong ${\bf q}$ dependence is caused by the RIXS scattering matrix elements. Whereas these same matrix elements forbid at ${\bf q}=0$ single- and bi-magnon scattering in first order, they act in the opposite way for three-magnon scattering, strongly enhancing its spectral weight at ${\bf q}=0$.

A proper computation of for instance the tri-magnon boundstate in a two-dimensional antiferromagnet would require on top of this calculation the inclusion of three-spin magnon-magnon scattering\cite{Millet1974,Himbergen1976,Cyr1996}, which is beyond our present scope. Considering magnon-magnon interactions in general terms, however, one expects these to reduce the three-magnon excitation energy, in analogy to two-magnon Raman scattering~\cite{Nagao2007,Canali1992,Chubukov1995}. The range of this energy reduction can be estimated by considering a N\'eel state of Ising spins: if three flipped spins form a string, the energy cost for such an excitation is $4J$, whereas one isolated spin-flip and two neighboring flipped spins cost $5J$, three isolated ones $6J$. Kinetics of course opposes three-magnon binding but in any case one expects a red-shift of tri-magnon spectral weight up to $2J$. 

\begin{figure}
\begin{center}
\includegraphics[width=.7\columnwidth,clip,trim=4cm 1cm 1cm 2.5cm]{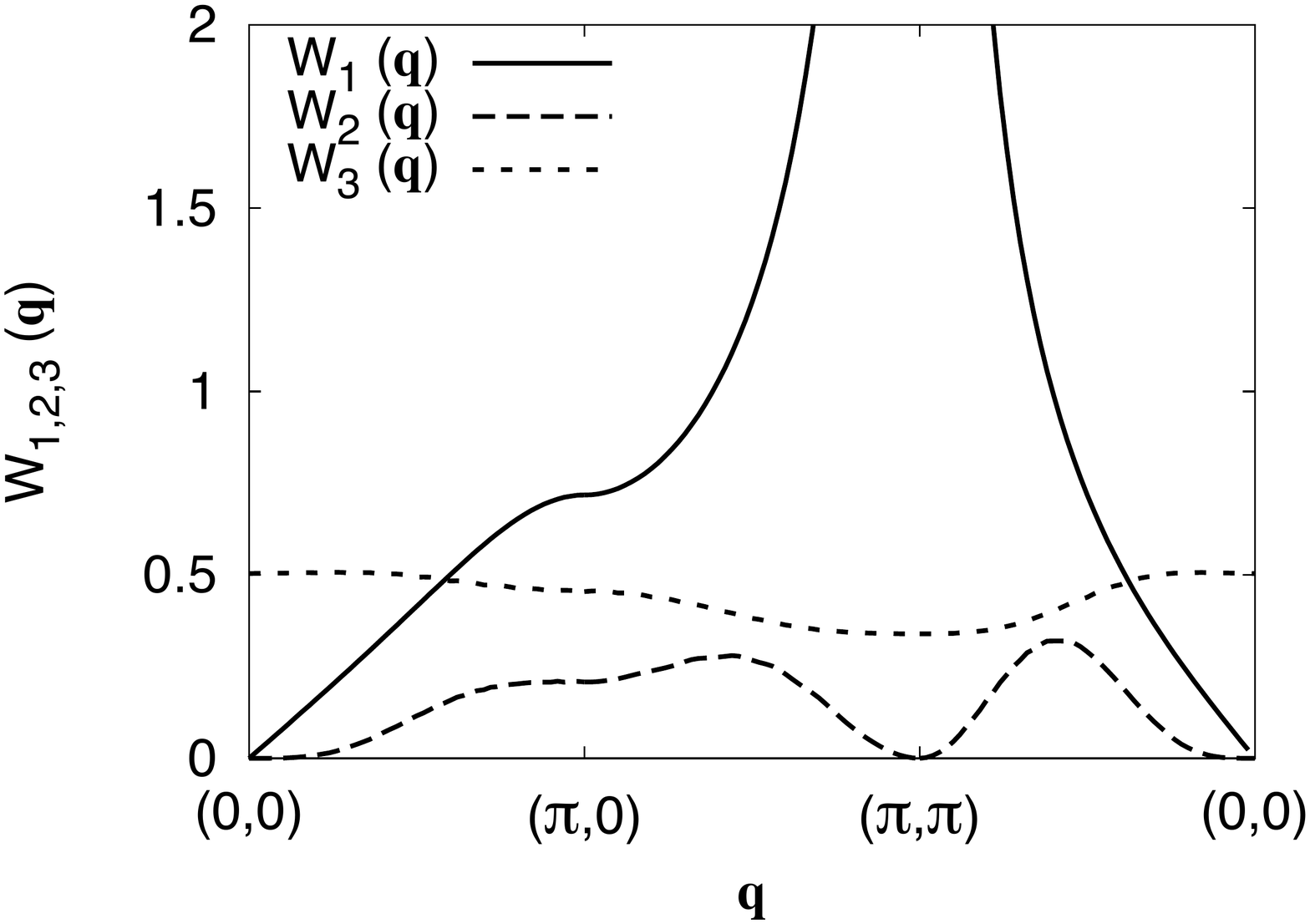}
\caption{Energy-integrated response functions $W_i ({\bf q})$ for single-, two- and three-magnon excitations ($i=1,2,3$, respectively). The ratio of the single- to three-magnon RIXS spectral weight is equal to $(\Gamma/J)^2 W_1 ({\bf q})/W_3 ({\bf q})$, see main text.}
\label{fig:W}
\end{center}
\end{figure}

{\it Three-magnon polarization dependence.} To unambiguously identify three-magnon spectral features in cuprates, knowledge of their theoretical polarization and scattering angle dependence is paramount. In evaluating the polarization factor $G({\boldsymbol \epsilon},{\boldsymbol \epsilon}')$, one is aided by the fact that in cuprates the magnetic moments are oriented in the CuO$_2$ planes\cite{Vaknin1987,Vaknin1989,Skanthakumar1989}.  Writing $G({\boldsymbol \epsilon},{\boldsymbol \epsilon}') = \epsilon'^*_i  G_{ij}^{\phantom{\dag}} \epsilon^{\phantom{\dag}}_j$ in Euclidian coordinates, the polarization factor for the Cu $L_{2,3}$- and $M_{2,3}$-edges simplifies to
\begin{equation}
  \left[ G_{ij} \right] \propto \begin{pmatrix} 0 & i & 0 \\ -i & 0 & 0 \\ 0 & 0 & 0 \end{pmatrix}.
\end{equation}
If in an experiment the particular polarization of the outgoing photon is not detected, the cross section is summed over the two outgoing polarizations ${\boldsymbol \epsilon}'$. When the $[0,\pi]$-direction (in the CuO$_2$ plane) is perpendicular to the scattering plane, and the incoming photons are $\pi$-polarized, one obtains\cite{Ament2009} $\sum_{{\boldsymbol \epsilon}'} G({\boldsymbol \epsilon},{\boldsymbol \epsilon}')^2 \propto \cos^2 \phi $, where $\phi$ is the angle of the incident beam with the normal to the CuO$_2$ planes. This holds for all scattering angles $2\theta$.

For the $M$-edge, changing the polarization vectors hardly affects ${\bf q}$ and therefore the spectral line shape is independent of the experimental geometry. It only affects the overall intensity, through $G({\boldsymbol \epsilon},{\boldsymbol \epsilon}')^2$. The Cu $L$-edge has a much higher energy: there is enough momentum available to probe approximately the entire BZ. Fig.~\ref{fig:Ledge} shows the three-magnon RIXS spectra for the case that the $[0,\pi]$-direction is perpendicular to the scattering plane with scattering angles $2\theta= 90^{\circ}$ and $130^{\circ}$ respectively. The asymmetry between $+q$ and $-q$ comes from the  $\cos^2 \phi$ polarization factor.

{\it Comparing one-, two- and three-magnon spectral weights.} Because both excitations have the same polarization factor, the ratio of three-magnon to single-magnon weight is independent of the polarization factor $G({\boldsymbol  \epsilon},{\boldsymbol  \epsilon}')$ within the UCL expansion. The RIXS intensity ratio is thus proportional to the ratio of the energy-integrated response functions of the two excitations, where the single-magnon response is given by\cite{Ament2009} $\chi_1 ({\bf q},\omega) \approx 0.718 (u_{\bf q}-v_{\bf q})^2 \delta (\omega - \omega_{\bf q})$. The numerical factor ensures the sum rule $ (1/N) \sum_{\bf q} \int d\omega\ \chi_1 ({\bf q},\omega) = 1$ is obeyed, a physical property of the Heisenberg model that needs to be enforced within linear spin wave theory. The ratio of the RIXS single-magnon to the three-magnon spectral weight is equal to $(\Gamma/J)^2 W_1({\bf q})/W_3 ({\bf q})$, where $W_i({\bf q}) = \int d\omega\ \chi_i ({\bf q},\omega)$. For typical cuprates at the Cu $M$-edge $J/\Gamma \approx 0.1$ and at the $L$-edge\cite{note1} $\approx 0.2-0.4$. Fig.~\ref{fig:W} shows the $W_i ({\bf q})$ of single-, two- and three-magnon excitations. $W_3 ({\bf q})$ decreases by $8 \%$ from $(0,0)$ to $(\pi,0)$, and by $30 \%$ from $(0,0)$ to $(\pi,\pi)$. The bimagnon RIXS spectral weight scales with the polarization factor for non-flip scattering,\cite{Ament2009} so a detailed comparison of it to the single- and three-magnon intensity involves the experimental geometry of choice. In any case we can conclude from Fig.~\ref{fig:W} that at ${\bf q}={\bf 0}$ and in a significant portion of the BZ around it the three-magnon spectral weight dominates over the other ones.

{\it Conclusions.} We have theoretically demonstrated the importance of the three-magnon scattering channel in RIXS at the Cu $L$- and $M$-edge of the high-$T_c$ cuprates. For small transferred momenta, ${\bf q} \rightarrow {\bf 0}$, three-magnon scattering dominates the magnetic RIXS spectrum. This scattering channel gives direct access to the dispersion of for instance tri-magnon boundstates in gapped low-dimensional quantum magnets. Experimentally there is evidence for tri-magnon spectral weight in Cu $M$-edge RIXS data\cite{Freelon2008} on CaCuO$_2$ showing a single peak feature at an energy loss of $450$ meV, consistent with our theory. This consistency can further be tested by measuring the polarization dependence and compare it to the theoretical one presented here. Also in very recent\cite{Braicovich_pc} Cu $L$-edge data on La$_2$CuO$_4$ the three-magnon feature appears to be present around $400$ meV for ${\bf q} =  {\bf 0}$. These results illustrate that RIXS has an unprecedented capacity to probe multi-spin responses in strongly correlated magnets -- both in principle and in practice.



\begin{thebibliography}{99}
\bibitem{Schuelke2007} W. Sch{\"u}lke, Electron Dynamics by Inelastic {X}-Ray Scattering,  Oxford University Press, 2007.
\bibitem{Kotani2001} A. Kotani and S. Shin, Rev. Mod. Phys. {\bf 73}, 203 (2001).
\bibitem{Hill2008} J.P. Hill {\it et al.}, Phys. Rev. Lett. {\bf 100}, 097001 (2008). 
\bibitem{Braicovich2009a} L. Braicovich {\it et al.}, Phys. Rev. Lett. {\bf 102}, 167401 (2009).
\bibitem{Schlappa2009} J. Schlappa  {\it et al.},  Phys. Rev. Lett. {\bf 103}, 047401, 2009.
\bibitem{Freelon2008} B. Freelon {\it et al.}, arXiv:0806.4432 (2008).
\bibitem{Braicovich2009b} L. Braicovich {\it et al.}, arXiv:0911.0621 (2009).
\bibitem{Ghiringhelli2009} G. Ghiringhelli {\it et al.}, Phys. Rev. Lett. {\bf 102}, 027401 (2009).
\bibitem{Brink2007} J. van den Brink, Europhys. Lett. {\bf 80}, 47003 (2007).
\bibitem{Nagao2007} T. Nagao and J. Igarashi, Phys. Rev. B {\bf 75}, 214414 (2007).
\bibitem{Forte2008} F. Forte, L.J.P. Ament and J. van den Brink, Phys. Rev. B {\bf 77}, 134428 (2008).
\bibitem{Ament2009} L.J.P. Ament {\it et al.}, Phys. Rev. Lett. {\bf 103}, 117003 (2009).
\bibitem{Haverkort2009} M. Haverkort, arXiv:0911.0706 (2009).
\bibitem{Brink2006}  J. van den Brink and M. van Veenendaal, Europhys. Lett., {\bf 73}, 121 (2006).
\bibitem{Devereaux2007}  T.P. Devereaux and R. Hackl, Rev. Mod. Phys. {\bf 79},  175  (2007).
\bibitem{Els1997}  G. Els  {\it et al.},   Phys. Rev. Lett. {\bf 79}, 5138 (1997).
\bibitem{Millet1974} P.J. Millet and H. Kaplan, Phys. Rev. B,  {\bf 10}, 3923 (1974).
\bibitem{Himbergen1976} J.E. van Himbergen and J.A. Tjon, Physica  {\bf 82A}, 389 (1976). 
\bibitem{Cyr1996} S.L.M. Cyr, B.W. Southern and D.A. Lavis, J. Phys.: Condens. Matter {\bf 8}, 4781 (1996).
\bibitem{Brink2005} J. van den Brink and M. van Veenendaal, J. Phys. Chem. Solids {\bf 66}, 2145 (2005).  
\bibitem{Ament2007} L.J.P. Ament, F. Forte, and J. van den Brink, Phys. Rev. B {\bf 75}, 115118 (2007).
\bibitem{Yin1974} L.I. Yin {\it et al.}, Phys. Rev. A {\bf 9}, 1070 (1974).
\bibitem{Coldea2001} R. Coldea {\it et al.},  Phys. Rev. Lett. {\bf 86}, 5377 (2001).
\bibitem{Vaknin1987} D. Vaknin {\it et al.}, Phys. Rev. Lett. {\bf 58}, 2802 (1987).
\bibitem{Vaknin1989} D. Vaknin {\it et al.},  Phys. Rev. B {\bf  39}, 9122 (1989).
\bibitem{Skanthakumar1989} S. Skanthakumar {\it et al.}, Physica C (Amsterdam) {\bf 160}, 124  (1989).
\bibitem{Yin1973} L.I. Yin, I. Adler, M.H. Chen, and B. Crasemann, Phys. Rev. A {\bf 7}, 897 (1973).
\bibitem{Canali1992} C.M. Canali, S.M. Girvin, Phys. Rev. B {\bf 45}, 7127 (1992).
\bibitem{Chubukov1995}  A.V. Chubukov and D.M. Frenkel, Phys. Rev. Lett. {\bf 74}, 3057 (1995).
\bibitem{note1} If $J/\Gamma$ is not small (e.g. $\sim 0.4$ at the Cu $L_3$-edge) then terms beyond the first order UCL become relevant. In this case one expects that three-magnon weight starts to compete with the single-magnon signal in a wider portion of the BZ. With respect to the first order UCL form shown in Fig.~\ref{fig:Ledge}, the three-magnon spectral weight will be redistributed to a certain extent. Moreover, if the intermediate state lives longer dynamical correlation functions that include more distant neighbors to the core-hole site increase in spectral weight.
\bibitem{Braicovich_pc} L. Braicovich, private communication.
\end{thebibliography}
\end{document}